\documentclass[aps,pre,twocolumn,groupedaddress,showpacs]{revtex4-1}

\usepackage{graphicx}
\usepackage{subfigure}
\usepackage{bm}
\begin{document}
\title{Sheared stably stratified turbulence
and large-scale waves in a lid driven cavity}
\author{N. Cohen}
\email{conim@post.bgu.ac.il}
\author{A. Eidelman}
\email{eidel@bgu.ac.il}
\author{T. Elperin}
\email{elperin@bgu.ac.il}
\homepage{http://www.bgu.ac.il/me/staff/tov}
\author{N. Kleeorin}
\email{nat@bgu.ac.il}
\author{I. Rogachevskii}
\email{gary@bgu.ac.il} \homepage{http://www.bgu.ac.il/~gary}
\affiliation{The Pearlstone Center for Aeronautical Engineering
Studies, Department of Mechanical Engineering, Ben-Gurion University
of the Negev, P.O.Box 653, Beer-Sheva 84105,
Israel}
\date{\today}
\begin{abstract}
We investigated experimentally stably stratified
turbulent flows in a lid driven cavity with a
non-zero vertical mean temperature gradient in
order to identify the parameters governing the
mean and turbulent flows and to understand their
effects on the momentum and heat transfer. We
found that the mean velocity patterns (e.g., the
form and the sizes of the large-scale
circulations) depend strongly on the degree of
the temperature stratification. In the case of
strong stable stratification, the strong
turbulence region is located in the vicinity of
the main large-scale circulation. We detected the
large-scale nonlinear oscillations in the case of
strong stable stratification which can be
interpreted as nonlinear internal gravity waves.
The ratio of the main energy-containing
frequencies of these waves in velocity and
temperature fields in the nonlinear stage is
about 2. The amplitude of the waves increases in
the region of weak turbulence (near the bottom
wall of the cavity), whereby the vertical mean
temperature gradient increases.
\end{abstract}


\maketitle

\section{Introduction}

A number of studies of turbulent transport in
lid-driven cavity flow have been conducted in the
past, because the lid-driven cavity is
encountered in many practical engineering and
industrial applications, and serves as a
benchmark problem for numerical simulations.
Detailed discussions of the state of the art of
the different studies of isothermal and
temperature stratified lid-driven turbulent
cavity flows have been published in several
reviews (see, e.g. Refs.~\onlinecite{SD00,KI08}).

Fluid flow and heat transfer in rectangular
cavities driven by buoyancy and shear have been
studied numerically and experimentally in a
number of publications (see, e.g.,
Refs.~\onlinecite{TD72,SK83,KSR84,KS84,KSA84,KSB84,KS85,KSN85,KP89,II89,MJ92,KS93,IHK93,MAV94,MV94,MV95,IH95,PK96,LC96,AR99,SMN02,OD04,LFG11}).
In particular, three-dimensional laminar
lid-driven cavity flow have been studied
experimentally and numerically (see
Refs.~\onlinecite{KS85,KSN85,KS93}), whereby the
Taylor-Gortler like (TGL) longitudinal vortices,
as well as other general flow structures, have
been found. In the isothermal flow, both the
number of vortex pairs and their average size
increases as the Reynolds number increases in
spite of the lateral confinement of the flow.
Different effects in a mixed convection in a lid
driven cavity have been investigated in the past
(see,
Refs.~\onlinecite{MJ92,MV94,PK96,LC96,SMN02,OD04}).
Excitation of an instability in a lid-driven flow
was studied in Ref.~\onlinecite{II89,LFG11}
numerically and experimentally. In addition,
lid-driven flow in a cube filled with a tap water
have been experimentally investigated in
Ref.~\onlinecite{LFG11} to validate the numerical
prediction of steady-oscillatory transition at
lower than ever observed Reynolds number.
The authors of Ref.~\onlinecite{LFG11}
reported that their experimental observations
agree with the numerical simulations that demonstrated the existence
of the large amplitude oscillatory motions in a lid-driven cavity flow.

There are several studies on lid driven cavity
flow with stable stratification.
Results of comprehensive experimental studies of flow patterns
and mixing in a stably-stratified, lid-driven water cavity flow
were reported in Ref.~\onlinecite{KSN85}.
These experiments were conducted in the range of Reynolds number from $10^3$  to $10^4$
and in the range of the bulk Richardson numbers from $0.08$
to  $6.5$. Stable stratification was controlled by varying
temperatures of the upper and bottom walls of the cavity.
The authors observed a strongly three dimensional flow with
TGL vortices which was partially turbulent.
Experiments revealed the existence of the primary and secondary
circulation cells with alternating sign of vorticity.

Three-dimensional numerical simulation in a
shallow driven cavity have been conducted for a
stably stratified fluid heated from the top
moving wall and cooled from below for a wide
range of Rayleigh numbers and Richardson numbers
(see Ref.~\onlinecite{MV95}). It was found that
an increase of the buoyancy force prevents the
return flow from penetrating to the bottom of the
cavity. The fluid is recirculated at the upper
portion of the cavity, and the upper
recirculation induces shear on the lower fluid
layer and forms another weak recirculated flow
region. Multicellular flow becomes evident when
the Richardson number is larger than $1$ and may
produce waves that propagate along the transverse
direction. Strong secondary circulation, and
separated flow are evident for the Richardson
number is about $0.1$. The rate of the heat
transfer increases as the Richardson number
decreases.

The two-dimensional and three-dimensional
numerical simulations in a driven cavity have
been conducted for a stably stratified fluid
heated from the top moving wall over broad ranges
of the parameters (see
Refs.~\onlinecite{IHK93,IH95}). It was found that
when the Richardson number is very small, the
gross flow characteristics are akin to the
conventional driven-cavity flows, as addressed by
earlier studies. In this case the isotherm
surfaces maintain a fair degree of
two-dimensionality. When the Richardson number
increases, the primary and meridional flows are
confined to the upper region of the cavity. In
the middle and lower parts of the cavity, fluid
tends to be stagnant, and heat transfer is mostly
conductive.

There are only a few experimental studies on lid
driven turbulent cavity flow with stable
stratification. In particular, stably-stratified
flows in a three-dimensional lid-driven cavity
have been experimentally studied in order to
examine the behavior of longitudinal
Taylor-Gortler-like vortices (see
Refs.~\onlinecite{KSR84,KS84,KSA84,KSB84,KI08}).
It was found that the Taylor-Gortler vortices
appear in all situations and they are generated
in the region of concave curvature of the flow
above a surface of separation in the shear flow.
In the stably stratified flow, the vortices
appear to enhance the mixing as they convolute
the interface. In the unstably stratified flow,
where the forced and free convection effects are
approximately in balance, the Taylor-Gortler
vortices are still formed. However, there are a
few Taylor-Gortler vortex pairs and their size is
small. The Taylor-Gortler vortices arise in the
region of the downstream secondary eddy and
corner vortices along the end-walls. At higher
Reynolds numbers $(\sim 10^4)$ the flow is
unsteady in the region of the downstream
secondary eddy and exhibits some turbulent
properties.

Stably stratified sheared turbulent flows are of
a great importance in atmospheric physics. Since
Richardson (1920), it was generally believed that
in stationary homogeneous atmospheric flows the
velocity shear becomes incapable of maintaining
turbulence when the Richardson number exceeds
some critical value (see, e.g.,
Refs.~\onlinecite{CHA61,MY71,MIL86}). The latter
assertion, however, contradicts to atmospheric
measurements, experimental evidence and numerical
simulations (see, e.g.,
Refs.~\onlinecite{SF01,BAN02,PAR02,MO02,LA04,M14}).
Recently an insight into this long-standing
problem has been gained through more rigorous
analysis of the turbulent energetics involving
additional budget equations for the turbulent
potential energy and turbulent heat flux, and
accounting for the energy exchange between
turbulent kinetic energy and turbulent potential
energy (see,
Refs.~\onlinecite{ZKR07,ZKR08,ZKR09,ZKR13}). This
analysis opens new prospects toward developing
consistent and practically useful turbulent
closures for stably stratified sheared turbulent
flows.

The main goal of this study is to investigate
experimentally stably stratified turbulent flows
in a lid driven cavity in order to identify the
parameters governing the mean and turbulent flows
and to understand their effects on the momentum
and heat transfer. This paper is organized as
follows. Section II describes the experimental
set-up and instrumentation. The results of
laboratory study of the stably stratified sheared
turbulent flow and comparison with the
theoretical predictions are described in Section
III. Finally, conclusions are drawn in Section
IV.

\section{Experimental set-up}

\begin{figure}
\vspace*{1mm} \centering
\includegraphics[width=7.5cm]{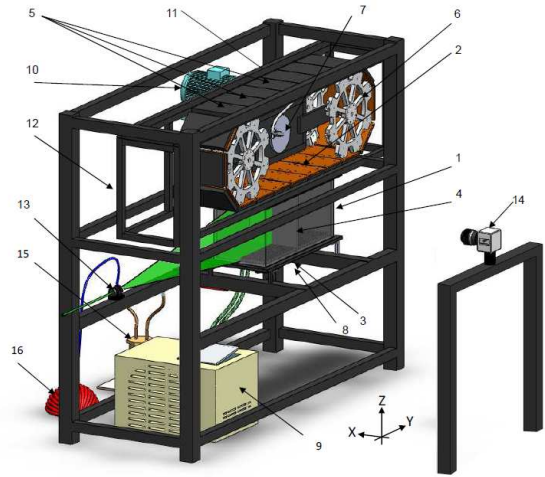}
\caption{\label{Fig1} Scheme of the experimental
set-up with sheared temperature stratified
turbulence: 1-Rectangular cavity; 2-Heated top
wall; 3-Cooled bottom wall; 4-Internal partition;
5-Plate heating elements; 6-Gear wheels;
7-Sliding rings; 8-Tank with cold water;
9-Chiller; 10-Electric motor; 11-Gear box;
12-Rigid steel frame; 13-Laser light sheet
optics;  14-CCD camera; 15-Generator of incense
smoke; 16-Pump.}
\end{figure}

The experiments have been carried out in a
lid-driven turbulent cavity flow generated by a
moving wall in rectangular cavity filled with air
(see Fig.~\ref{Fig1}). An internal partition
which is parallel to the $XZ$-plane is inserted
in the cavity in order to vary the aspect ratio
of the cavity. Here we introduce the following
system of coordinates: $Z$ is the vertical axis,
the $Y$-axis is a direction of a long wall and
the $XZ$-plane is parallel to a square side wall
of the cavity. A top wall of the cavity moves in
a $Y$-axis direction and generates a shear flow
in the cavity. The experiments have been
conducted in the cavity with the dimensions $26.4
\times 29 \times 27.4$ cm$^3$. Heated top wall and
cooled bottom wall of the cavity impose a
temperature gradient in the flow which causes
temperature stratification of the air inside the
cavity.

The top moving wall of the chamber consists of
identical rectangular plate heating elements with
a width of $16$ cm which are connected by hinges
to the two adjustment heating elements. Twenty
connected heating elements form a closed conveyer
belt which is driven by two rotating gear-wheels
with hinges. At each moment 6 moving heating
elements are located in the plane and form the
moving top wall of the cavity. Each heating
element comprises the aluminum plate with the
attached electrical heater and a temperature
probe and is insulated with a textolite cap. The
heaters and the temperature probes are
electrically connected to a power supply unit and
to the measuring device through a set of sliding
rings. A bottom stationary cold wall of the
cavity is manufactured from aluminum and serves
as a top wall of a tank filled with water which
circulates through a chiller with a controlled
temperature.

All moving parts of the experimental set-up
including an electrical motor and a gear-box are
attached to a rigid steel frame in order to
minimize vibrations of the cavity that is
connected to the moving wall with a soft flexible
sealing. Perspex walls are attached to the frame
and enclose the moving conveyer belt consisting
of heating elements in order to reduce heat
transfer from the heating elements. The
experiments have been conducted at different
velocities of the top wall and aspect ratios of
the cavity, and at different temperature
differences between the top and the bottom walls
of the cavity. This experimental set-up allows us
to produce sheared temperature stratified
turbulence.

The turbulent velocity field have been measured
using a digital Particle Image Velocimetry (PIV)
system (see, e.g.,
Refs.~\onlinecite{AD91,W00,RWK07}) with LaVision
Flow Master III. A double-pulsed Nd-YAG laser
(Continuum Surelite $ 2 \times 170$ mJ) is used
for light sheet formation. Light sheet optics
comprise spherical and cylindrical Galilei
telescopes with tuneable divergence and
adjustable focus length. We employ a
progressive-scan 12 Bit digital CCD camera
(pixels with a size $6.7 \,\mu$m $\times 6.7 \,
\mu$m each) with dual frame technique for
cross-correlation processing of captured images.
The tracer used for PIV measurements is incense
smoke with sub-micron particles (with the
material density $ \rho_{\rm tr} \approx 1 $
g/cm$^3$), which is produced by high temperature
sublimation of solid incense particles. Velocity
measurements were conducted in two perpendicular
cross-sections in the cavity, $YZ$ cross-section
(frontal view) and $XZ$ cross-section (side
view).

We have determined the mean and the
r.m.s.~velocities, two-point correlation
functions and an integral scale of turbulence
from the measured velocity fields. Series of 520
pairs of images, acquired with a frequency of 1
Hz, have been stored for calculating velocity
maps and for ensemble and spatial averaging of
turbulence characteristics. The center of the
measurement region coincides with the center of
the chamber. We have measured velocity in the
probed cross-section $240 \times 290$ mm$^2$ with
a spatial resolution of $2048 \times 2048$
pixels. This corresponds to a spatial resolution
142 $\mu$m / pixel. This probed region has been
analyzed with interrogation windows of $32 \times
32$ or $16 \times 16$ pixels, respectively.

In every interrogation window a velocity vector
have been determined from which velocity maps
comprising $32 \times 32$ or $64 \times 64$
vectors are constructed.
The mean and r.m.s. velocities for every point of a
velocity map (1024 or 4096 points) have been calculated by
averaging over 520 independent maps. Space averaging was used
for determining the mean energy in several regions in the flow and
calculating the mean characteristic velocities in these regions
for different stratifications.
The two-point correlation functions of the velocity field have
been calculated for every point of the velocity
map inside the main vortex (with $32 \times 32$
vectors) by averaging over 520 independent
velocity maps, and then they will be averaged
over all points. An integral scale $\ell_0$ of
turbulence has been determined from the two-point
correlation functions of the velocity field.

The temperature field has been measured with a
temperature probe equipped with twelve
E-thermocouples (with the diameter of 0.13 mm and
the sensitivity of $65 \, \mu$V/K) attached to a
vertical rod with a diameter 4 mm. The spacing
between thermocouples along the rod is 22 mm.
Each thermocouple is inserted into a 1 mm
diameter and 45 mm long case. A tip of a
thermocouple protrudes at the length of 15 mm out
of the case. The mean temperature is measured for
10 rod positions with 25 mm intervals in the
horizontal direction, i.e., at 120 locations in a
flow. The exact position of each thermocouple is
measured using images captured with the optical
system employed in PIV measurements. A sequence
of temperature readings (each reading is averaged
over 50 instantaneous measurements which are
obtained in 20 ms) for every thermocouple at
every rod position is recorded and processed
using the developed software based on LabVIEW
7.0. The measurements from 12 thermocouples are
obtained every 0.8 s. Similar experimental
technique and data processing procedure were used
previously in the experimental study of different
aspects of turbulent convection, stably
stratified turbulent flows (see
Refs.~\onlinecite{BEKR11,EEKR06,EKR13}) and in
Refs.~\onlinecite{BEE04,EE04,EEKR06C,EEKR06A,EKR10}
for investigating a phenomenon of turbulent
thermal diffusion (see
Refs.~\onlinecite{EKR96,EKR97}).

\section{Experimental results and comparison with the theoretical predictions}

\begin{figure}
\vspace*{1mm} \centering
\includegraphics[width=7.5cm]{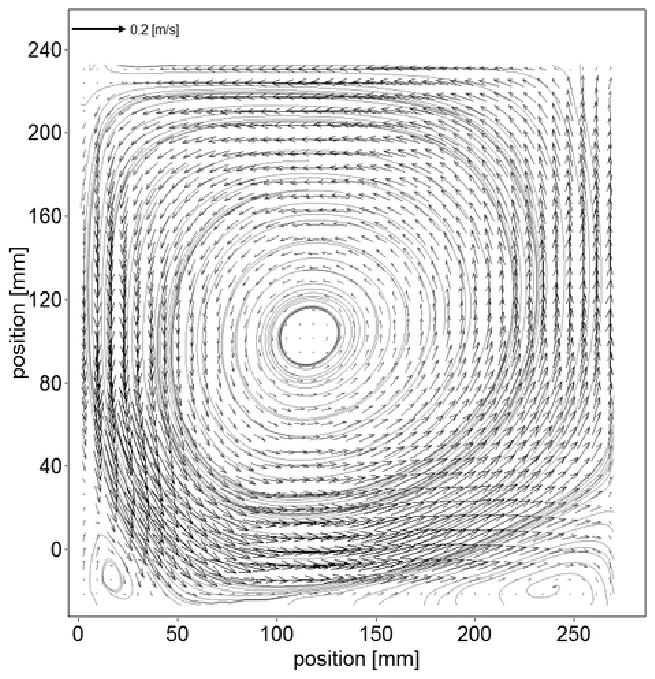}
\includegraphics[width=7.5cm]{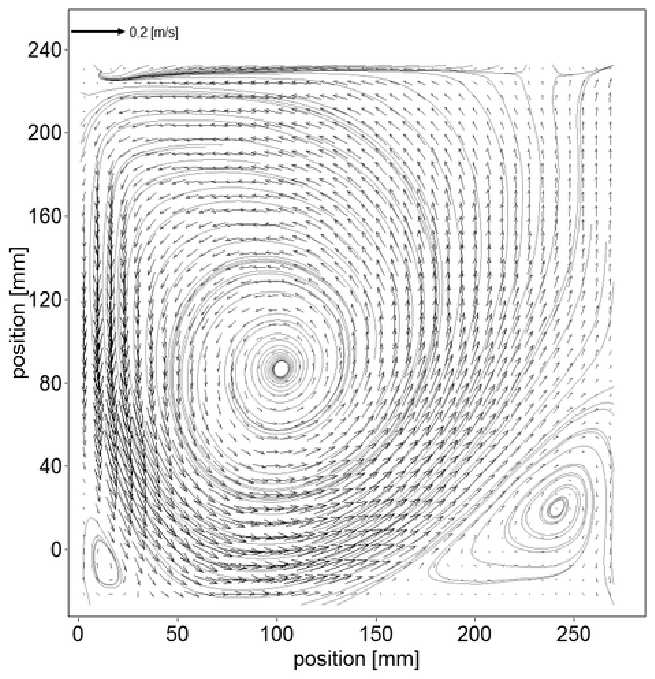}
\includegraphics[width=7.5cm]{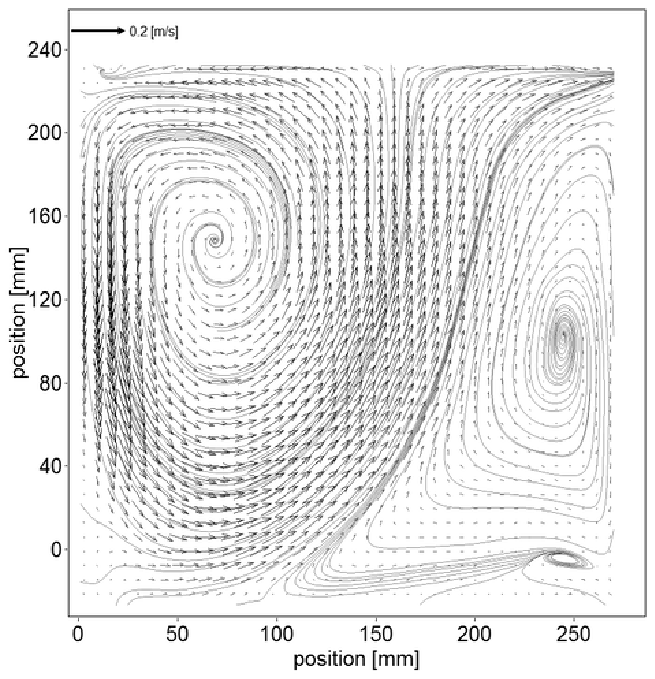}
\caption{\label{Fig2} Mean flow patterns obtained
in the experiments in $YZ$ cross-section at the
different temperature differences between the top
and bottom walls: $\Delta T = 0$ K (upper panel);
$\Delta T =11$ K (middle panel); $\Delta T = 21$
K (lower panel). Coordinates $y$ and $z$ are
measured in mm.}
\end{figure}

\begin{figure}
\vspace*{1mm} \centering
\includegraphics[width=7.5cm]{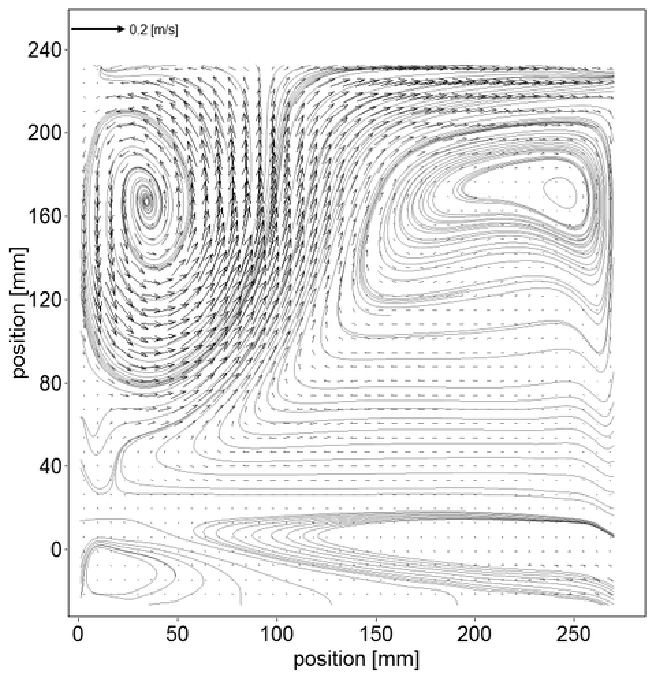}
\includegraphics[width=7.5cm]{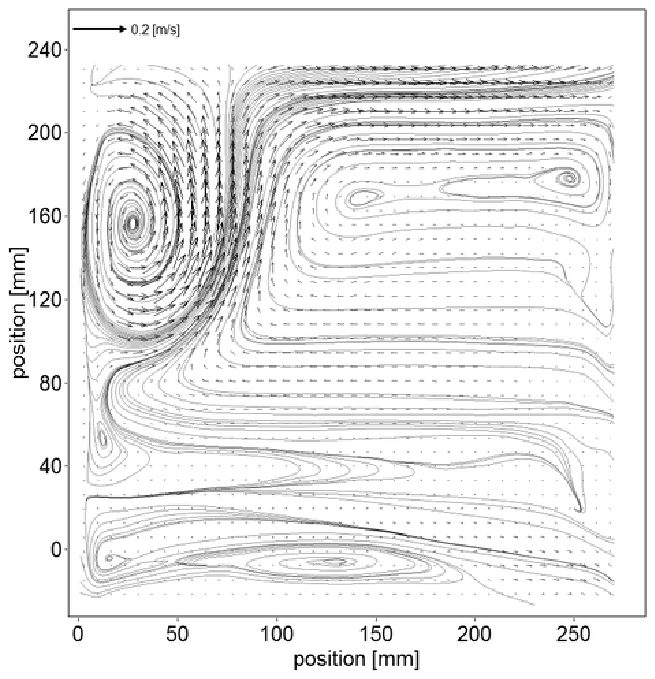}
\includegraphics[width=7.5cm]{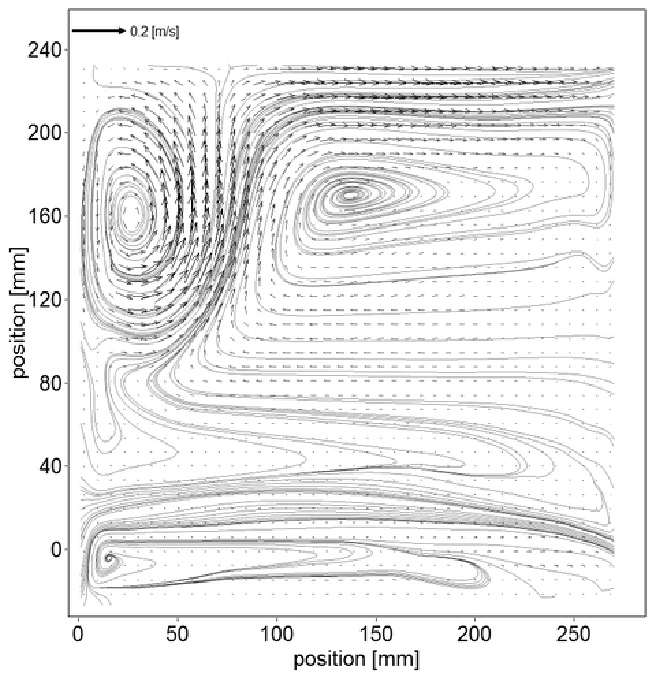}
\caption{\label{Fig3} Mean flow patterns obtained
in the experiments in $YZ$ cross-section at the
different temperature differences between the top
and bottom walls: $\Delta T = 33$ K (upper
panel); $\Delta T =44$ K (middle panel); $\Delta
T = 54$ K (lower panel). Coordinates $y$ and $z$
are measured in mm.}
\end{figure}

We start the analysis of the experimental results
with the mean flow patterns obtained in the
experiments conducted at different values of the
temperature difference $\Delta T$ between the top
and bottom walls. A set of mean velocity fields
obtained in the central $YZ$ plane is shown in
Figs.~\ref{Fig2} and~\ref{Fig3}. The portion of flow
depicted in these figures is far from the lid. These
experiments demonstrate strong modification of
the mean flow patterns with an increasing
temperature difference $\Delta T$ between the hot
top wall and the cold bottom wall of the cavity
whereby the top wall moves in the left direction.

Figures~\ref{Fig2} and~\ref{Fig3} demonstrate the
major qualitative changes in the mean flow
patterns as the bulk Richardson number, ${\rm
Ri}_b= g \alpha \Delta T H_z / U_0^2$,
encompasses a wide range. Here $\alpha$ is the
thermal expansion coefficient, $U_0 = 118$ cm/s
is the lid velocity, $H_z = 24$ cm is the
vertical height of the cavity and $g$ is the
gravitational acceleration. The primary mean
circulation (the main vortex), shown in the upper
panel of Fig.~\ref{Fig2}, occupies the entire
cavity. Two weak secondary mean vortexes are
observed at the lower corners of the cavity. The
qualitative character of the mean flow for small
$\Delta T$ (or ${\rm Ri}_b \ll 1)$ is similar to
the conventional lid driven cavity flow of a
non-stratified fluid.

When $\Delta T$ (or ${\rm Ri}_b)$ is gradually
increased, a change of the mean flow is observed
already at a relatively low temperature
difference [compare the middle panel of
Fig.~\ref{Fig2} that corresponds to $\Delta T=11$
K (or ${\rm Ri}_b=0.06)$, with the bottom panel
of Fig.~\ref{Fig2} that is for $\Delta T=21$ K,
${\rm Ri}_b=0.12$]. The position of the main
vortex is shifted to the left, while the position
of the right weak secondary vortex is shifted
upwards and its size increases. At a further
increase of the temperature difference $\Delta T$
between the top and bottom walls, $33 \leq \Delta
T \leq 54$ K $(0.19 \leq {\rm Ri}_b \leq 0.29)$,
the main vortex is pushed upwards, its size
decreases, and the weak secondary mean flow is
also strongly changed (see Fig.~\ref{Fig3}).

\begin{figure}
\vspace*{1mm} \centering
\includegraphics[width=7.5cm]{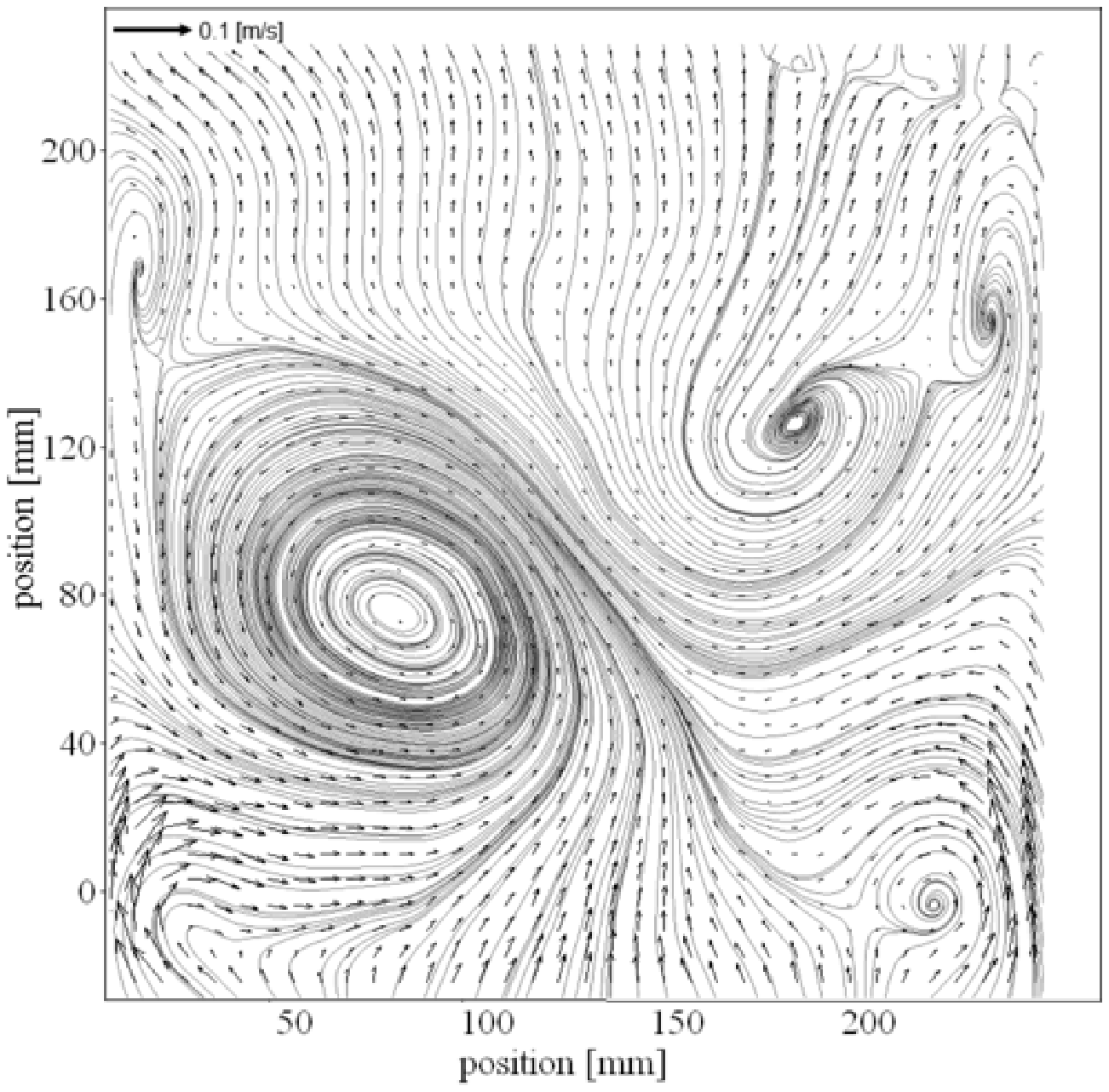}
\includegraphics[width=7.5cm]{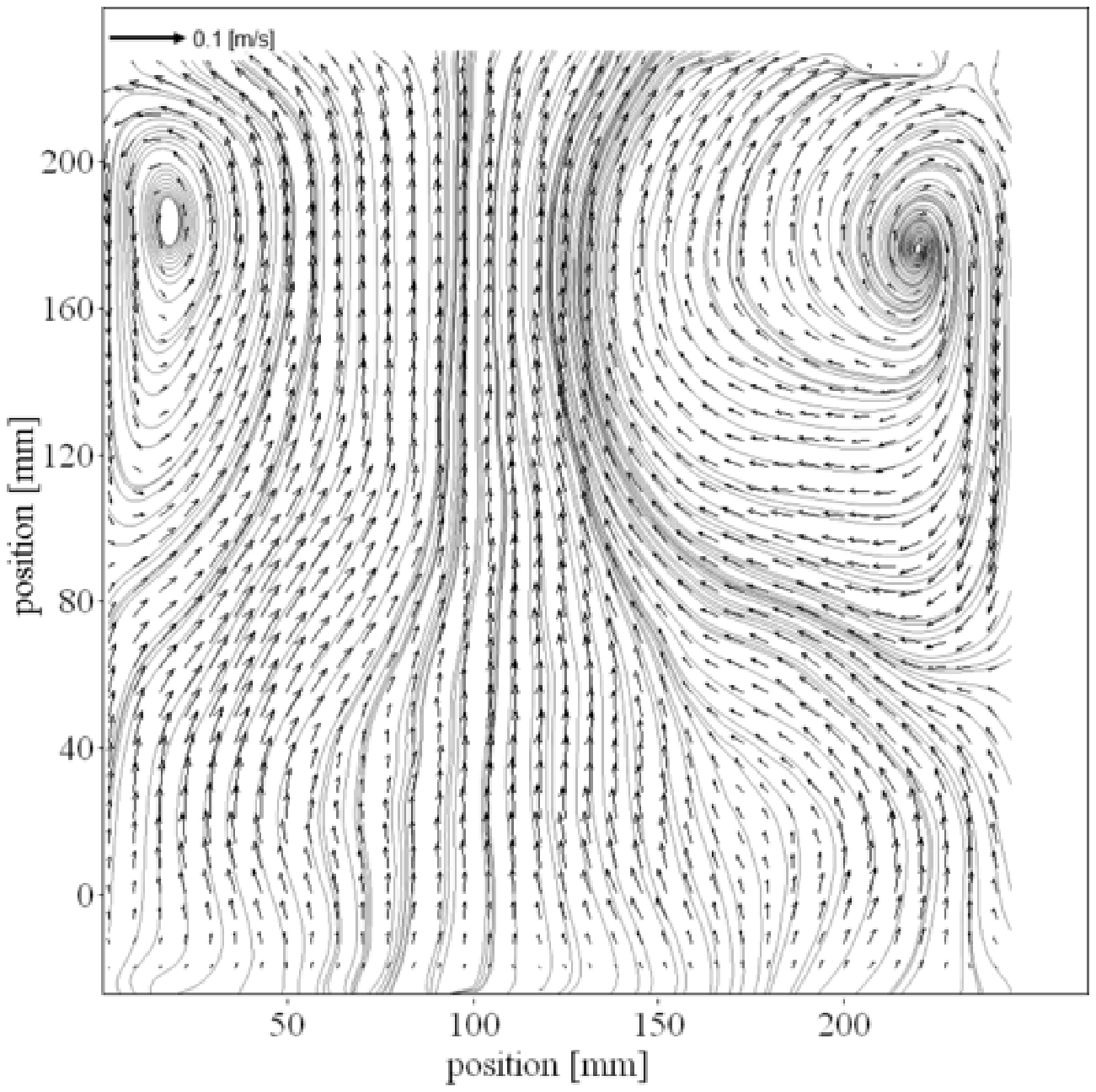}
\includegraphics[width=7.5cm]{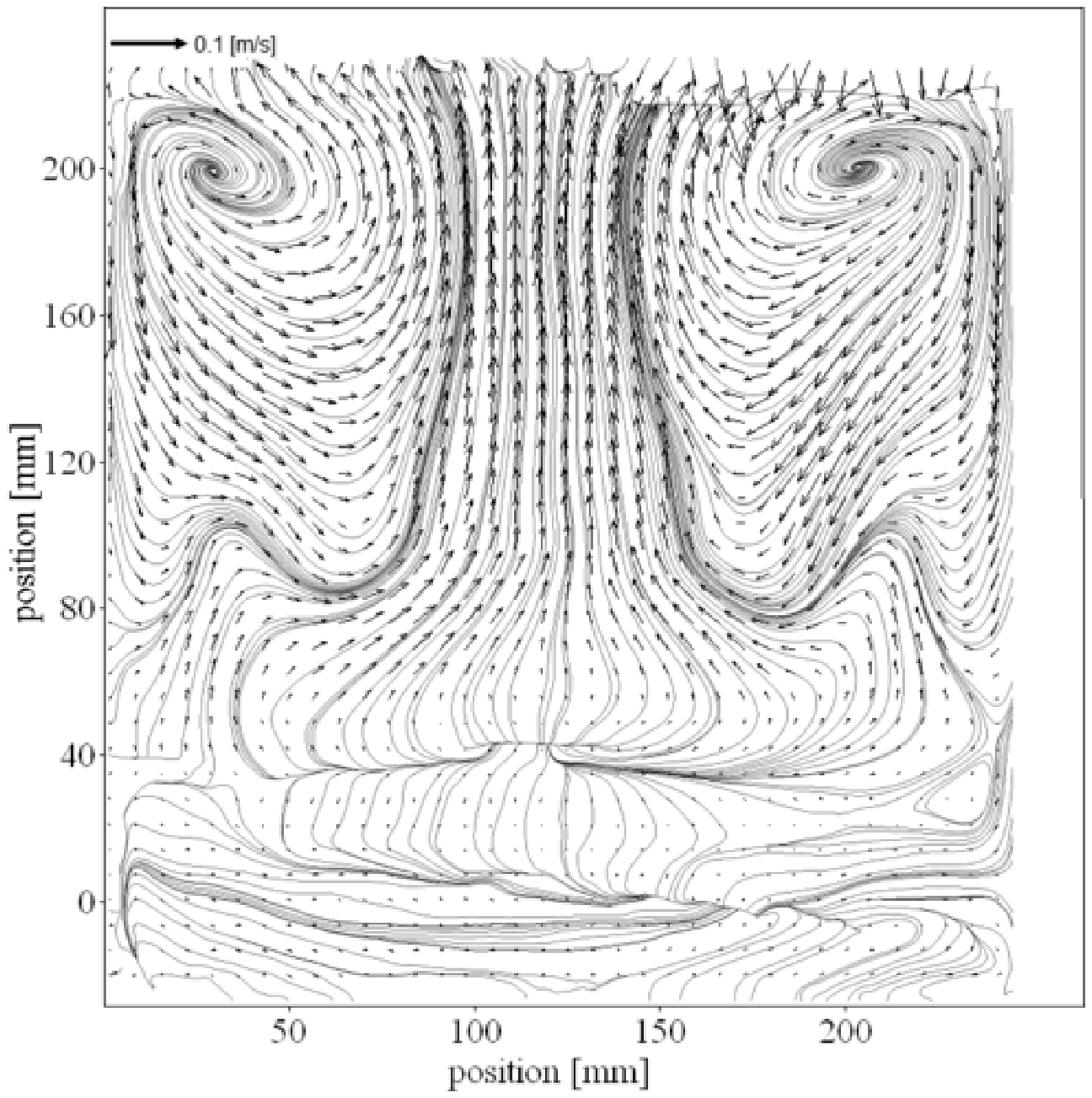}
\caption{\label{Fig4} Mean flow patterns
obtained in the experiments in $XZ$ cross-section
at the different temperature differences between
the top and bottom walls: $\Delta T = 0$ K at $y=12$ cm (upper
panel); $\Delta T =21$ K at $y=8$ cm (middle panel); $\Delta
T = 40$ K at $y=4$ cm (lower panel). Coordinates $x$ and $z$
are measured in mm.}
\end{figure}

Strong effect of stratification on the flow
pattern in the cavity can be also seen by
inspecting mean velocity maps in $XZ$
cross-section (see Fig.~\ref{Fig4}) obtained for different temperature
difference between the top and bottom walls. The
mean velocity fields shown in Fig.~\ref{Fig4}
are measured in the cross-section close to the
center of the main vortex. The frontal and side
velocity maps demonstrate a complex
three-dimensional flow in the lid-driven cavity
which comprises the main vortex seen in frontal
view, and several secondary vortices (seen in
frontal and side views).

It must be noted that the mean flow energy in
the central part of the $XZ$ cross-section is
by at least one order of magnitude less than the mean flow energy in the
front $YZ$ cross-section. There are several factors
that can cause a slight asymmetry of the mean
flow in $XZ$ plane, in particular for $\Delta T =0$. One
of the factors can be related with slight geometric
imperfections of the experimental cell. Another
possible reason is associated with the effect
of excitation of the large-scale vorticity in a non-stratified
turbulence with a large-scale shear (see Refs.~\onlinecite{EKR03,EKR07}).
Temperature stratification may
mitigate this effect. The latter assertion is supported
by the mean velocity maps in Fig.~\ref{Fig4} (see
lower panel, $\Delta T =40$ K) where the large-scale
velocity pattern in $XZ$ plane is quite symmetric.

It is possible to distinguish between three
regions in the mean flow for $\Delta T \geq 33$ K
or ${\rm Ri}_b \geq 0.19$ (see the upper panel of
Fig.~\ref{Fig3}): a relatively strong mean flow
in the upper part of the cavity including a main
vortex in its left side, a mean sheared flow with
a lesser mean velocity in its right side, and a
very weak mean flow in the rest of the cavity.
Therefore, when the strength of stable
stratification increases and the bulk Richardson
number, ${\rm Ri}_b$, increases up to 0.3, the
main vortex tends to be confined to a small zone
close to the sliding top lid and to the left wall
of the cavity.

The stable stratification suppresses the vertical
mean motions, and, therefore, the impact of the
sliding top wall penetrates to the smaller depth
into the fluid. As seen in Fig.~\ref{Fig3}, when
${\rm Ri}_b$ is not small, the mean flow in the
middle and lower parts of the cavity interior is
weak, and much of the fluid remains almost
stagnant. The lower panel of Fig.~\ref{Fig3}
demonstrates this trend. The mean flow is almost
stagnant in the bulk of the cavity interior
excluding the region close to the sliding top
wall.

The mean flows obtained in our experiment are
three-dimensional. However, inspection of the
mean flow patterns in $YZ$ (Figs.~\ref{Fig2} and~\ref{Fig3}) and
$XZ$ planes (Fig.~\ref{Fig4}) reveals that the energy of the
mean flow in $YZ$ plane is by at least one order
of magnitude larger than the mean flow energy
in $XZ$ plane. Therefore, the flow is not strongly
three-dimensional with respect to the mean flow
energy in these two planes. Moreover, inspection
of the velocity maps in $XZ$ cross-section (see
Fig.~\ref{Fig4}, panels for $\Delta T =21$ K and $\Delta T =40$ K) shows
that the flow is nearly symmetric and homogeneous
in the central part. We also observed that
the measured velocity maps in $YZ$ cross-sections
for different coordinate $x$ are quite close, and a
slight difference was observed only at the distance
of about 3.7 cm from the side walls. Therefore,
we believe that the choice of the mid plane provides
a good representation of the core flow in the
cavity.

\begin{figure}
\vspace*{1mm} \centering
\includegraphics[width=7.5cm]{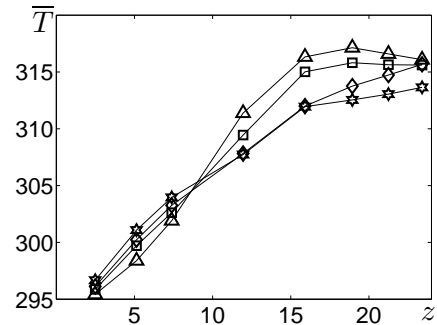}
\caption{\label{Fig14}
Spatial vertical profiles of mean temperature field $\overline{T}(z)$
in the central $YZ$ cross-section for different $y$: 2.5 cm
(triangles), 10 cm (squares), 14 cm (diamonds), 22 cm (stars),
and at the temperature difference $\Delta T=$ 44 K between the
bottom and the top walls of the chamber obtained in the
experiments. The mean temperature $\overline{T}$ is measured in K,
while the lengths $z$ is measured in cm.}
\end{figure}

In Fig.~\ref{Fig14} we show vertical temperature
profiles measured in the central $YZ$ plane for several
values of coordinate $y$. Inspection
of these temperature profiles reveals the existence
of two distinctive parts in the temperature
dependence on vertical coordinate z. Temperature
varies strongly in the bottom stagnant part
of the flow with a stable stratification. Vertical
temperature gradient varies from 1.7 K/cm (at $y$ = 2.5 cm) under
the main vortex to 1.15 K/cm (at $y$ = 14 cm and $y$ = 22 cm) under a
sheared part of the flow (tail of the vortex). Vertical temperature gradient
decreases sharply to 0.2 K/cm inside the sheared
part of the flow. Remarkably, vertical temperature
gradient changes sign inside the main vortex
(see the plots in Fig.~\ref{Fig14} for $y$ = 2.5 cm), i.e. stably
stratified flow changes to the slightly unstably
stratified flow.

\begin{figure}
\vspace*{1mm} \centering
\includegraphics[width=7.5cm]{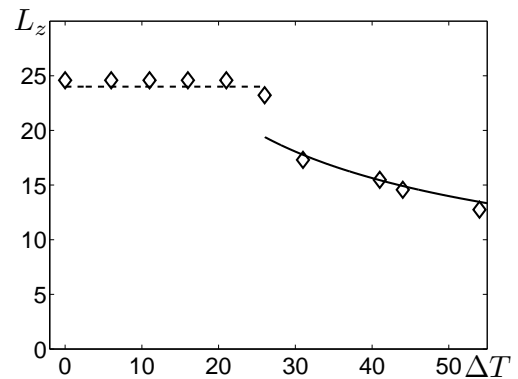}
\caption{\label{Fig5} Maximum vertical size $L_z$
of large-scale circulation versus the temperature
difference $\Delta T$ between the bottom and the
top  walls of the chamber obtained in the
experiments. Dashed lines correspond to  fitting
of the experimental points, while solid line
corresponds to the theoretical estimates. The
temperature difference $\Delta T$ is measured in
K, while the size $L_z$ is measured in cm.}
\end{figure}

To characterize the change in the mean flow
pattern with the increase of the stratification,
we show in Fig.~\ref{Fig5} the maximum vertical
size $L_z$ of the main (energy containing)
large-scale circulation versus the temperature
difference $\Delta T$ between the bottom and the
top  walls of the chamber obtained in the
experiments.
The length $L_z$, the vertical size of the large-scale vortex, is determined
directly from the measured mean flow energy values.
The vertical coordinate of the bottom of the vortex was determined
as a location in the flow where the mean flow energy is by a factor
10  smaller than the maximum mean flow energy.
For example, $L_z$ = 14.6 cm for $\Delta T$=44 K and $L_z$ = 12.9 cm for $\Delta T$=54 K.
Inspection of Fig.~\ref{Fig5} shows
that the maximum vertical size $L_z$ of
large-scale circulation is nearly constant when
$\Delta T < 25$ K. On the other hand, when
$\Delta T > 25$ K, the maximum vertical size
$L_z$ decreases with $\Delta T$ as $L_z \propto
1/\Delta T$.

\begin{figure}
\vspace*{1mm} \centering
\includegraphics[width=7.5cm]{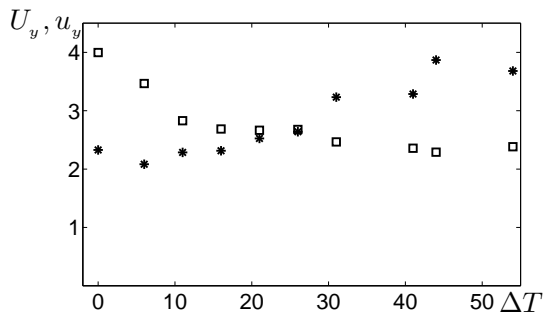}
\caption{\label{Fig6} Characteristic horizontal
mean $U_y$ (squares) and turbulent $u_y$
(snowflakes) velocities versus the temperature
difference $\Delta T$ between the bottom and the
top  walls of the chamber obtained in the
experiments. The temperature difference $\Delta
T$ is measured in K, while the velocities are
measured in cm/s.}
\end{figure}

The point of maximum penetration corresponds
to the region where the buoyancy force exerted
by the lower denser fluid balances the net downward
momentum flux of the fluid generated by the lid motion
(see Ref.~\onlinecite{KSN85}). Therefore, this scaling
can be understood on the base of the budget equations
for the mean velocity and
temperature fields. Indeed, using the budget
equation for the mean kinetic energy $E_U = \rho
{\bm U}^2/2$, we obtain that the change of the
mean kinetic energy $\delta E_U$ is of the order
of the work of the buoyancy force, $\rho {\bm
U}^2/2 \sim \rho \beta (\delta T) L_z$, where
$\beta=g/T_0$ is the buoyancy parameter, $T$ is
the mean temperature with the reference value
$T_0$, and $\delta T$ is the change of the mean
temperature over the size of the large-scale
circulation $L_z$. On the other hand, the budget
equation for the squared mean temperature, $T^2$,
shows that the change of the mean temperature
$\delta T$ over the size of the large-scale
circulation $L_z$ is of the order of the
temperature difference $\Delta T$ between the
bottom and the top  walls of the chamber. This
yields the following scaling:
\begin{eqnarray}
L_z \sim {\bm U}^2/ \beta \Delta T .
 \label{B1}
\end{eqnarray}

There are two regions in the flow where the
magnitude of turbulent kinetic energy differs significantly
for all stratifications with $\Delta T > 21$ K.
Strong (weak) turbulence is observed in regions
with strong (weak) mean flow. For instance, in
the case with the largest stratification $\Delta T > 54$
K obtained in our experiments, the turbulent kinetic
energy in the strong turbulence region is by
a factor 30 larger than that in the
weak turbulence region. The strong turbulence is
produced by the shear of large-scale vortex.

\begin{figure}
\vspace*{1mm} \centering
\includegraphics[width=7.5cm]{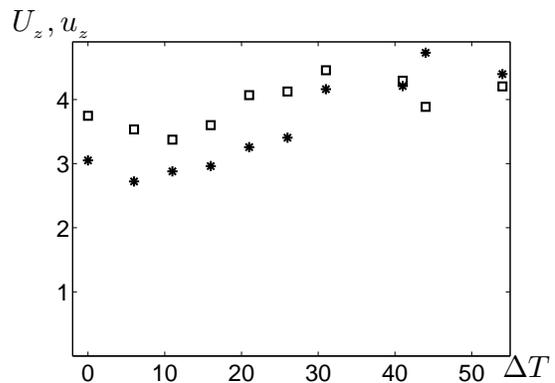}
\caption{\label{Fig7} Characteristic vertical
mean $U_z$ (squares) and turbulent $u_z$
(snowflakes) velocities versus the temperature
difference $\Delta T$ between the bottom and the
top  walls of the chamber obtained in the
experiments. The temperature difference $\Delta
T$ is measured in K, while the velocities are
measured in cm/s.}
\end{figure}

To characterize the velocity
fields obtained in the experiments, in
Figs.~\ref{Fig6} and~\ref{Fig7} we show the
characteristic horizontal and vertical mean,
$U_{y,z}$, and turbulent, $u_{y,z}$, r.m.s.
velocities (measured in the region with a strong
turbulence) versus the temperature difference
$\Delta T$ between the bottom and the top  walls
of the chamber. Inspection of Fig.~\ref{Fig6}
shows when $\Delta T < 25$ K, the horizontal mean
velocity, $U_{y}$, is larger than the turbulent
velocity $u_{y}$, while for $\Delta T > 25$ K,
$U_{y} < u_{y}$. This tendency is related to the
fact that for $\Delta T > 25$ K, the size of the
main (energy containing) large-scale circulation
decreases with $\Delta T$.

\begin{figure}
\vspace*{1mm} \centering
\includegraphics[width=7.5cm]{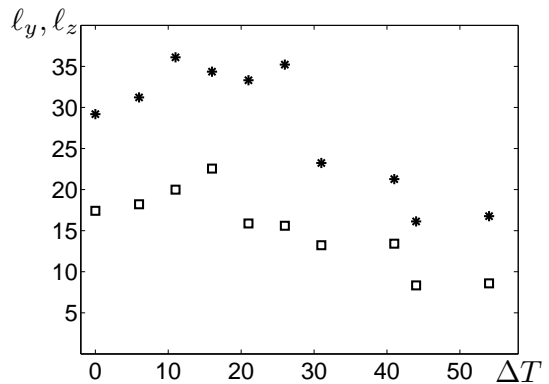}
\caption{\label{Fig8} Integral scales of
turbulence in horizontal $\ell_y$ (snowflakes)
and vertical $\ell_z$ (squares) directions versus
the temperature difference $\Delta T$ between the
bottom and the top  walls of the chamber obtained
in the experiments. The temperature difference
$\Delta T$ is measured in K, while the lengths
$\ell_{y,z}$ are measured in mm.}
\end{figure}

This implies that the size of the shear-produced
turbulence region due to the main vortex
decreases with increase of the stratification. In
this region, the mean velocity $U_y$ is nearly
constant. Substituting the vertical size of the
turbulence region, $L_z$, determined by
Eq.~(\ref{B1}) into the mean shear $S =dU_{y}/dz
\sim U_y/L_z$, we obtain
\begin{eqnarray}
{dU_{y} / dz} \sim U_y \beta \Delta T /{\bm U}^2
\sim \beta \Delta T / U,
 \label{B2}
\end{eqnarray}
where we have taken into account that $U_y \sim
U$. Therefore, the value of shear increases with
$\Delta T$, and, consequently, the shear
production rate $\Pi=\nu_{_{T}} S^2$ increases
with the increasing of the stratification, where
$\nu_{_{T}} \sim \ell_z u_z$ is the turbulent
viscosity and $\ell_z$ is the integral scale of
turbulence in the vertical direction. Now let us
estimate the turbulent kinetic energy, $\rho {\bm
u}^2/2$, using the budget equation for this
quantity:
\begin{eqnarray}
\rho {\bm u}^2/2 \sim \Pi \ell_z /u_z \sim
\ell_z^2 S^2 \sim \ell_z^2 (\beta \Delta T /
U)^2,
 \label{B3}
\end{eqnarray}
where ${\bm u}$ is the turbulent r.m.s. velocity.
Since the mean velocity is nearly constant for
$\Delta T > 25$ K, and $u \propto \ell_z \Delta
T$ [see Eq.~(\ref{B3})], the turbulent velocity
increases with the increase of the
stratification. Here we have taken into account
that the integral scale of turbulence in vertical
direction does not change strongly with the
change of $\Delta T$ when $\Delta T > 25$ K (see
Fig.~\ref{Fig8}).

\begin{figure}
\vspace*{1mm} \centering
\includegraphics[width=7.5cm]{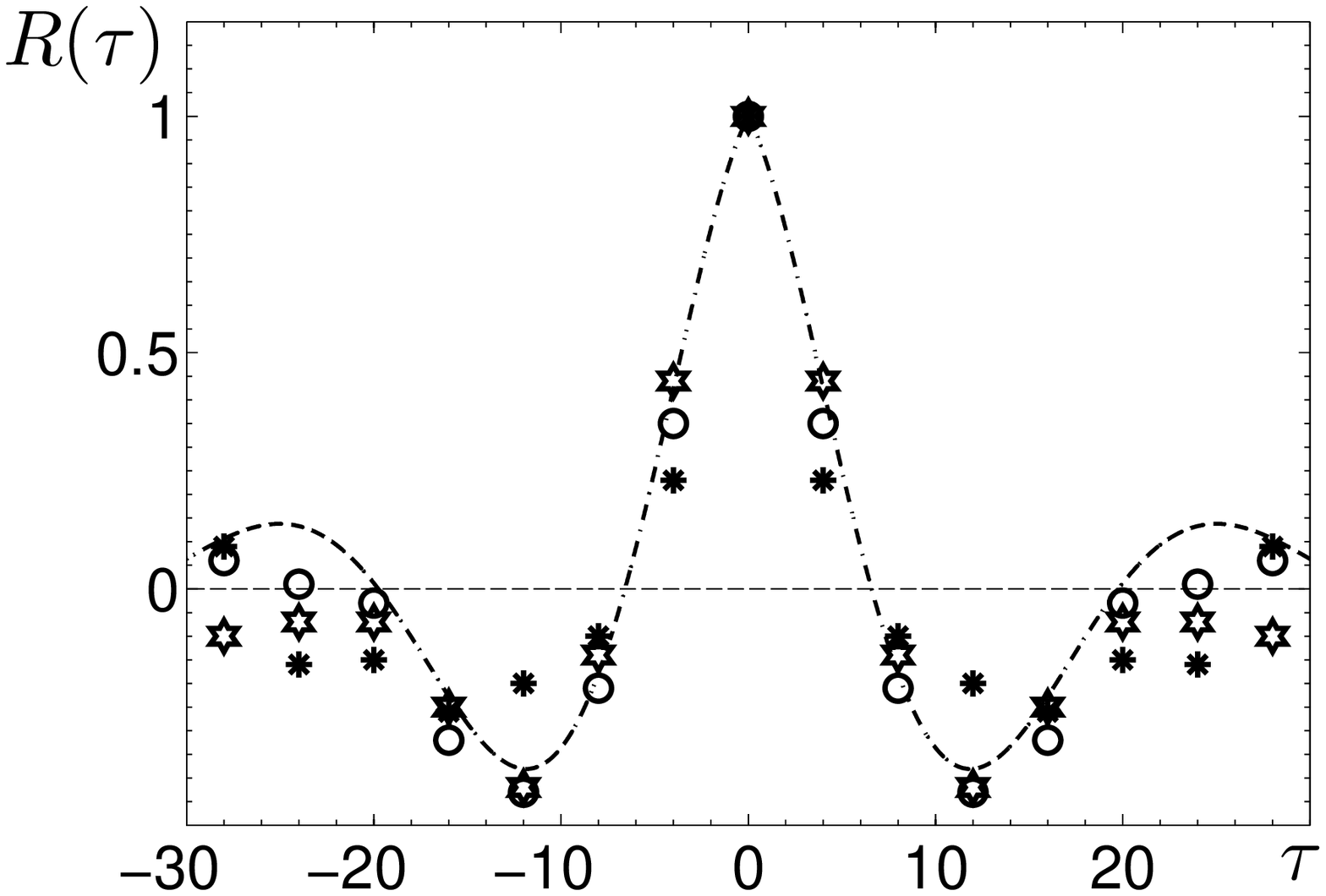}
\includegraphics[width=7.5cm]{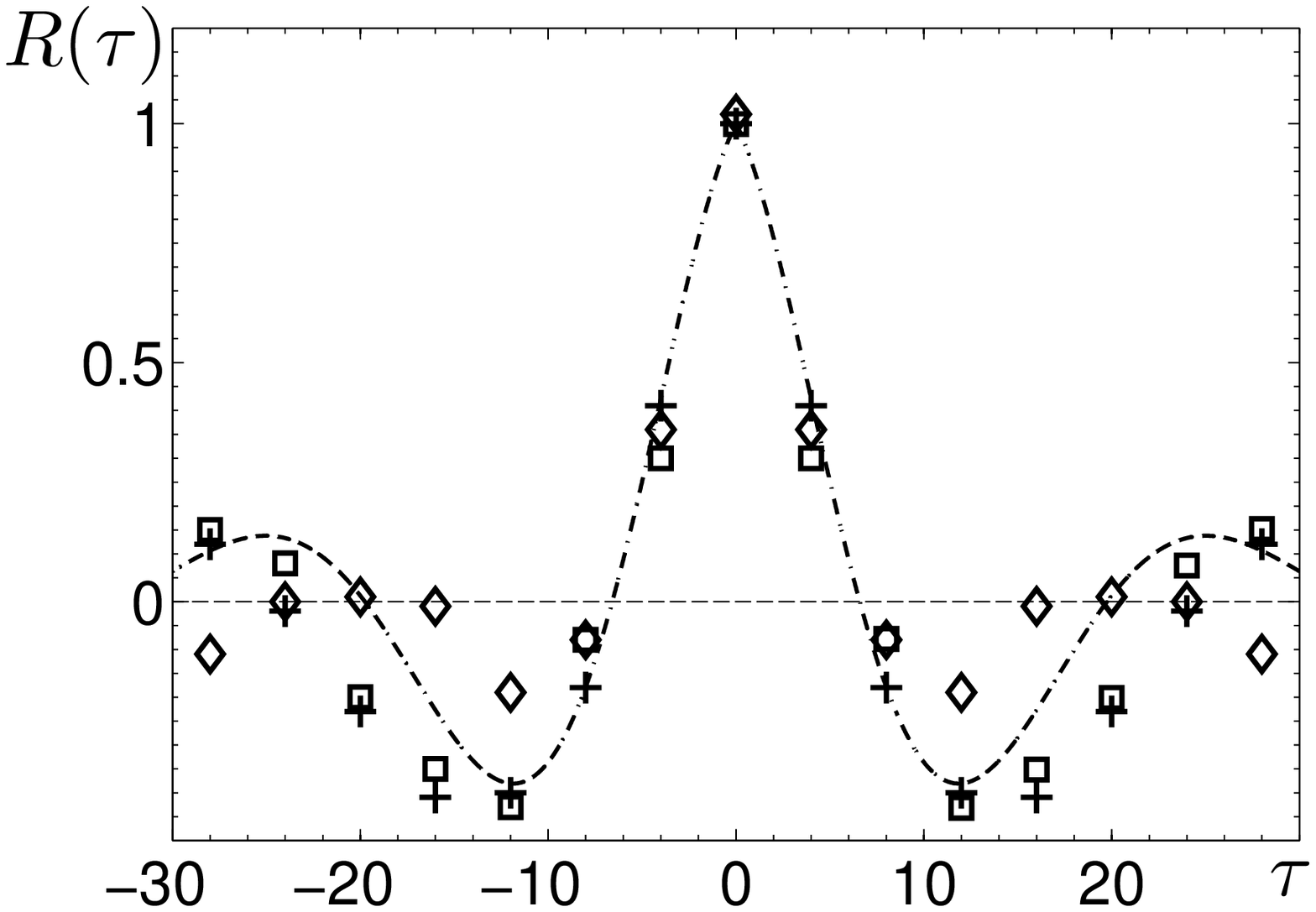}
\caption{\label{Fig9} Normalized one-point
non-instantaneous correlation function $R(\tau) =
\langle \delta T(z,t) \delta T(z,t+\tau) \rangle
/ \langle \delta T^2(z,t) \rangle$ of the
large-scale temperature field determined for
different $z$: 2.5 cm (diamonds), 5.1 cm
(six-pointed stars), 7.4 cm (crosses), 11.9 cm
(snowflakes), 15.9 cm (squares), 18.9 cm
(circles) shown in upper and lower panels, versus
the time $\tau$ at the temperature difference
$\Delta T=54$ K between the bottom and the top
walls of the chamber obtained in the experiments,
where $\delta T = \overline{T} - \overline{T}_0$.
The dashed fitting line corresponds to the
Lorentz function with $\tau_0=13$ s and
$\omega_0=0.286$ s$^{-1}$. The time
$\tau$ is measured in seconds.}
\end{figure}

The internal gravity waves with the frequency
$\omega = N k_h/k$ can be excited in stably
stratified flows. Here $k$ is the wave number,
$k_h$ is the horizontal wave number and $N=
(\beta \, \nabla_z T)^{1/2}$ is the
Brunt-V\"{a}is\"{a}l\"{a} frequency (see, e.g.
Refs.~\onlinecite{MY71,M14,ZKR09,M01,SS02,F88}).
In our experiments in the region of the cavity
with a weak turbulence we observed the
large-scale internal gravity waves with the
period of about 22 seconds. In particular, in
Fig.~\ref{Fig9} we show the normalized one-point
non-instantaneous correlation function $R(\tau) =
\langle \delta T(z,t) \delta T(z,t+\tau) \rangle
/ \langle \delta T^2(z,t) \rangle$ of the
large-scale temperature field determined for
different $z$ versus the time $\tau$, where
$\delta T = \overline{T} - \overline{T}_0$, and
$\overline{T}$ is the sliding averaged
temperature (with $10$ seconds window average),
$\overline{T}_0=\langle\overline{T}\rangle^{(sa)}$
and $\langle ... \rangle^{(sa)}$ is the 10
minutes average.

Inspection of Fig.~\ref{Fig9}
shows that the function $R(\tau)$ has a form of
the Lorentz function, $R(\tau)
=\exp(-\tau/\tau_0) \cos(\omega_0 \tau)$ with
$\tau_0=13$ s and $\omega_0=0.286$ s$^{-1}$,
which corresponds to the period of the wave
$2\pi/\omega_0=22$ seconds. Note that the Fourier
transform, $R(\omega)$, of the Lorentz function
has the following form:
\begin{eqnarray}
R(\omega)={2 \pi \over \tau_0} \Big(
[(\omega-\omega_0)^2 + \tau_0^{-2}]^{-1} +
[(\omega+\omega_0)^2 + \tau_0^{-2}]^{-1}\Big).
\nonumber\\
 \label{B4}
\end{eqnarray}
Such form of the correlation function $R(\tau)$
indicates the presence of the large-scale waves
with random phases. The memory or correlation
time for these waves is about $11$ s. Therefore,
in our analysis the temperature field is
decomposed in three different parts: small-scale
temperature fluctuations, the mean temperature
field and the large-scale temperature field
corresponding to the large-scale internal gravity
waves.

\begin{figure}
\vspace*{1mm} \centering
\includegraphics[width=7.5cm]{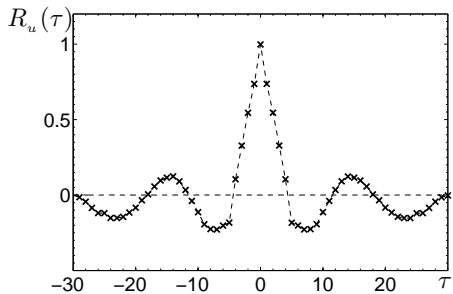}
\caption{\label{Fig10} Normalized one-point
non-instantaneous correlation function $R_u(\tau)
= \langle \delta U_z(z,t) \delta U_z(z,t+\tau)
\rangle / \langle \delta U_z^2(z,t) \rangle$ of
the vertical large-scale  velocity field
determined for different $z$ versus the time
$\tau$ determined for different $z$ versus the
time $\tau$ at the temperature difference $\Delta
T=54$ K between the bottom and the top  walls of
the chamber obtained in the experiments, where
$\delta U_z = \overline{U_z} - \overline{U_z}_0$.
The time $\tau$ is measured in seconds.}
\end{figure}

We also performed similar analysis for the
vertical large-scale velocity field. In our
analysis the velocity field is decomposed in
three different parts: small-scale velocity
fluctuations, the mean velocity field and the
large-scale velocity field corresponding to the
large-scale internal gravity waves.
Internal gravity waves are characterized
by $\delta U_z = \overline{U_z} - \overline{U}_{z0}$,
where $\overline{U_z}$ is the sliding averaged
vertical mean velocity (with 3 seconds
window average), $\overline{U}_{z0}=\langle \overline{U_z} \rangle^{(sa)}$,
and $\langle ... \rangle^{(sa)}$ is the 8.7 minutes average.
In Fig.~\ref{Fig10} we show the normalized one-point
non-instantaneous correlation function $R_u(\tau)
= \langle \delta U_z(z,t) \delta U_z(z,t+\tau)
\rangle / \langle \delta U_z^2(z,t) \rangle$ of
the vertical large-scale velocity field
determined for different $z$ versus the time
$\tau$.

\begin{figure}
\vspace*{1mm} \centering
\includegraphics[width=7.5cm]{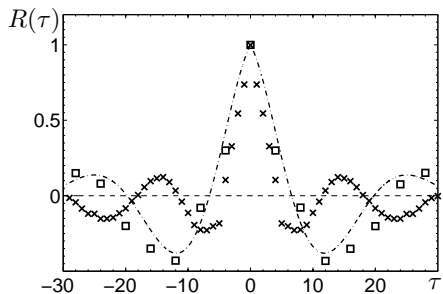}
\caption{\label{Fig11} Comparison of the
normalized one-point non-instantaneous
correlation function, $R_u(\tau)$ (crosses), of
the vertical large-scale velocity field with that
of the large-scale temperature field $R(\tau)$
(squares) determined for different $z$ versus the
time $\tau$ at the temperature difference $\Delta
T=54$ K between the bottom and the top  walls of
the chamber obtained in the experiments. The time
$\tau$ is measured in seconds.}
\end{figure}

Comparison of the normalized one-point
non-instantaneous correlation function,
$R_u(\tau)$, of the vertical large-scale velocity
field with that of the temperature field,
$R(\tau)$ (see Fig.~\ref{Fig11}) shows that for
short time scales ($\tau < 10$ s) these
correlation functions are different. This implies
that for these time-scales the wave spectra of
the large-scale velocity and temperature fields
are different.

\begin{figure}
\vspace*{1mm} \centering
\includegraphics[width=7.5cm]{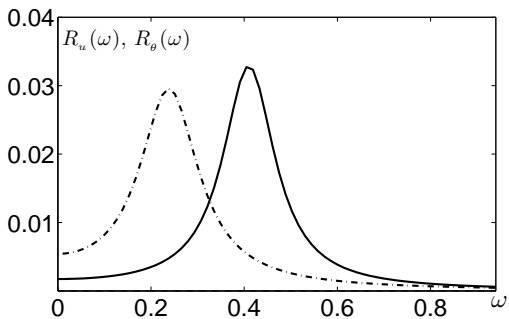}
\caption{\label{Fig12} The spectral functions,
$R_u(\omega)= (2 \pi)^{-1} \int R_u(\tau) \exp(-i
\omega \tau) \,d\tau$ (solid) and
$R_\theta(\omega)= (2 \pi)^{-1} \int R(\tau)
\exp(-i \omega \tau) \,d\tau$ (dashed-dotted), of
the normalized one-point non-instantaneous
correlation functions, $R_u(\tau)$ and $R(\tau)$,
of the vertical large-scale velocity and
temperature fields, where the function
$R_u(\tau)$ is shown in Figs.~\ref{Fig10}
and~\ref{Fig11}, while the function $R(\tau)$ is
shown in Fig.~\ref{Fig9}. The frequency $\omega$ is measured in
s$^{-1}$.}
\end{figure}

Indeed, in Fig.~\ref{Fig12} we show the spectral
functions, $R_u(\omega)= (2 \pi)^{-1} \int
R_u(\tau) \exp(-i \omega \tau) \,d\tau$ and
$R_\theta(\omega)= (2 \pi)^{-1} \int R(\tau)
\exp(-i \omega \tau) \,d\tau$, of the normalized
one-point non-instantaneous correlation
functions, $R_u(\tau)$ and $R(\tau)$, of the
vertical large-scale velocity and temperature
fields.
Figure~\ref{Fig12} shows a single peak for each variable: the
vertical large-scale velocity and the large-scale
temperature field, with 2:1 frequency ratio for
these fields. The reason for such behaviour
may be a parametric nonlinear excitation and
interaction of the wave temperature and velocity
fields. This also can be an interaction of nonlinear
internal gravity waves and the large-scale
Tollmien-Schlichting waves in sheared turbulent
flows (see Ref.~\onlinecite{EKR07}). The observed features can
be interpreted as a combination of standing
and propagating waves which can be excited
by the interaction of the mean flow
and the walls of the cavity, or by wave-radiation
stress mechanism pointed out in Refs.~\onlinecite{M01} and~\onlinecite{F88}.

\begin{figure}
\vspace*{1mm} \centering
\includegraphics[width=7.5cm]{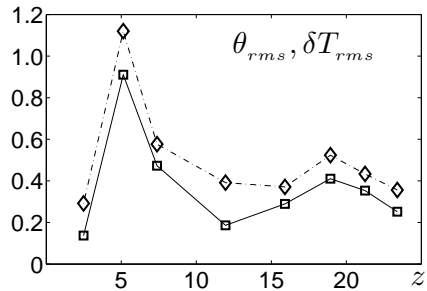}
\caption{\label{Fig13} Spatial vertical profiles
of turbulent temperature fluctuations
$\theta_{\it rms}$ (dashed) and of the function
$\delta T_{\it rms} \equiv \langle \delta T^2(z)
\rangle^{1/2}$ of the large-scale temperature
field (solid) at the temperature difference
$\Delta T=33$ K between the bottom and the top
walls of the chamber obtained in the experiments,
where $\delta T = \overline{T} - \overline{T}_0$.
The temperatures are measured
in K, while the lengths $z$ is measured in cm.}
\end{figure}

In Fig.~\ref{Fig13} we show spatial vertical
profiles of turbulent temperature fluctuations
$\theta_{\it rms}$ and of the function $\delta
T_{\it rms} \equiv \langle \delta T^2(z)
\rangle^{1/2}$ of the large-scale temperature
field at the temperature difference $\Delta T=33$
K between the bottom and the top walls of the
chamber obtained in the experiments, where
$\delta T = \overline{T} - \overline{T}_0$.
Inspection of Fig.~\ref{Fig13} shows that the
level of the intensities of turbulent temperature
fluctuations are of the same order as the energy
of the large-scale internal gravity waves. These
turbulent fluctuations, $\theta_{\it rms}$, are
larger in the lower part of the cavity where the
mean temperature gradient is maximum. In the
upper part of the cavity where the shear caused
by the large-scale circulation is maximum, and
the mean temperature gradient is decreased.

\section{Conclusions}

We study experimentally stably stratified
turbulence and large-scale flows and waves in a
lid driven cavity with a non-zero vertical mean
temperature gradient. Geometrical properties of
the large-scale vortex (e.g., its size and form)
and the level of small-scale turbulence inside
the vortex are controlled by the buoyancy (i.e.,
by the temperature stratification). The observed
velocity fluctuations are produced by the shear
of the large-scale vortex. At larger
stratification obtained in our experiments, the
strong turbulence region is located at the upper
left part of the cavity where the large-scale
vortex exists. In this region the
Brunt-V\"{a}is\"{a}l\"{a} frequency is small and
increases in the direction outside the
large-scale vortex. This is the reason of that
the large-scale internal gravity waves are
observed in the regions outside the large-scale
vortex. We found these waves by analyzing the
non-instantaneous correlation functions of the
temperature and velocity fields. The observed
large-scale waves are nonlinear because the
frequency of the waves determined from the
temperature field measurements is two times
smaller than that obtained from the velocity
field measurements. The measured intensity of the
waves is of the order of the level of the
temperature turbulent fluctuations.

\begin{acknowledgements}
We thank A.~Krein for his assistance in
construction of the experimental set-up. This
research was supported in part by the Israel
Science Foundation governed by the Israeli
Academy of Sciences (Grant 1037/11), and by the
Russian Government Mega Grant (Grant
11.G34.31.0048).
\end{acknowledgements}

\end{document}